\documentclass[aps,prl,reprint,showpacs,amsmath,superscriptaddress]{revtex4-1}
\usepackage{amssymb}
\usepackage{mathrsfs}
\usepackage{graphicx}
\usepackage{float}
\usepackage{txfonts}
\usepackage[normalem]{ulem}
\usepackage{color}

\begin{document}

\title{$2\pi$-flux loop semimetals}

\author{Linhu Li}
\affiliation{Beijing Computational Science Research Center, Beijing 100089, China}
\affiliation{CeFEMA, Instituto Superior
T\'ecnico, Universidade de Lisboa, Av. Rovisco Pais, 1049-001 Lisboa,  Portugal}
\author{Stefano Chesi}
\affiliation{Beijing Computational Science Research Center, Beijing 100089, China}
\author{Chuanhao Yin}
\affiliation{Beijing National Laboratory for Condensed Matter Physics, Institute of Physics, Chinese Academy of Sciences, Beijing 100190, China}
\affiliation{School of Physical Sciences, University of Chinese Academy of Sciences, Beijing 100049, China}
\author{Shu Chen}
\affiliation{Beijing National Laboratory for Condensed Matter Physics, Institute of Physics, Chinese Academy of Sciences, Beijing 100190, China}
\affiliation{School of Physical Sciences, University of Chinese Academy of Sciences, Beijing 100049, China}
\affiliation{Collaborative Innovation Center of Quantum Matter, Beijing, China}

\begin{abstract}
We introduce a model of $2\pi$-flux loop semimetals which holds nodal loops described by a winding number $\nu=2$. By adding some extra terms, this model can be transformed into a recently discovered Hopf-link semimetal, and the symmetries distinguishing these two phases are studied. We also propose a simpler physical implementation of $2\pi$-flux loops and of the Hopf-link semimetals which only involves nearest-neighbor hoppings, although in the presence of spin-orbit interaction.
Finally, we investigate the Floquet properties of the $2\pi$-flux loop, and find that such a loop may be driven into two separated $\pi$-flux loops or four Weyl points
by light with circular polarization in certain directions.
\end{abstract}

\maketitle
\emph{Introduction.-}Topological phases of matter have been one of the most intriguing research subjects in condensed matter physics. Beyond the topological insulators and superconductors characterized by a band gap and topological in-gap edge states\cite{Hasan_2010,Qi_2011,Shen_book,Bernevig_book}, and the nodal point semimetals with discrete gap-closing points in the Brillouin zone\cite{Wan_2011,Young_2012,Morimoto_2014,Yang_2014}, the recent discovery of topological nodal line semimetals (NLSMs) has opened a new chapter in the studies of topological phases. A NLSM has one or several one-dimensional (1D) gap-closing loops in momentum space, and the system holds flat-band surface states within the loops, namely the drumhead surface states. These surface states correspond to a nontrivial topological invariant of the bulk states, i.e. a $\pi$ Berry phase along a trajectory enclosing a nodal loop\cite{Burkov_2011,Weng_2015,Kim_2015,Yu_2015,Zhang_2016}.

Compared to the zero-dimensional (0D) gap-closing points in nodal point semimetals, the 1D nodes in NLSMs support much richer structures, and many efforts have been made on studying various aspects of NLSMs, e.g. nodal lines with a $Z_2$ monopole charge\cite{Fang_2015,Fang_2016}, nodal chain semimetals which hold nodal lines touching at some 0D points\cite{Bzdusek_2016,Yu_2017}, tunable Weyl points generated from NLSMs\cite{Yan_2016}, NLSMs with a nodal line acting as a vortex ring\cite{Lim_2017}, and the relation between NLSMs and topological insulators\cite{Li2016,Li2017}. More recently, a brand new type of NLSMs, the Hopf-link semimetals, has been reported by several different group independently\cite{Chen_Hopf,Yan_Hopf,Chang_Hopf,Ezawa_Hopf}. These semimetals have a band structure with a pair of nodal lines linked to each other, which is described by the Hopf maps and a topological invariant known as the Hopf number\cite{Wilczek_1983,Nakahara_2003}.

Despite these various efforts on investigating NLSMs, most of the loops being studied carry a Berry phase of $\pi$, and fall into a $Z_2$ classification. However, the Berry phase is associated to a winding number, which is a $Z$ invariant. This fact indicates that one may realize a system with nodal loops of high winding number, which corresponds to
vortices with multiples of $\pi$. In this letter, we construct a tight-binding model with winding number $\nu=2$, namely a $2\pi$-flux loop semimetal. By adding extra terms, the model can be transformed into a Hopf-link semimetal. The symmetries related to these two types of semimetals are also discussed. Then we introduce a cubic lattice model with spin-orbit coupling, which support a $2\pi$-flux gapless loop. Finally, we consider the continuum limit of our model, and find that the $2\pi$-flux loop can be decoupled into two $\pi$-flux loops
or two pairs of Weyl points,
under a periodic driving of a circularly polarized light (CPL).

\emph{$2\pi$-flux nodal loops and topological properties.-} We begin our discussion with a semimetallic model with conventional nodal loops, which can be described by the low energy effective Hamiltonian:
\begin{eqnarray}
h(\mathbf{k})=(k^2-m_0)\sigma_1+k_z\sigma_3,
\end{eqnarray}
with $k=\sqrt{k_x^2+k_y^2}$ and $\sigma_i$ the Pauli matrices. This model gives a nodal loop at $\mathbf{L}=(k,k_z)=(\sqrt{m_0},0)$, which is characterized by a Berry phase $\gamma=\pi$ along a trajectory enclosing the loop. For instance, By choosing a tiny circle enclosing the loop with a phase angle $\theta$, the expansion near the loop yields
\begin{eqnarray}
h(\mathbf{q})=2\sqrt{m_0}\cos{\theta}\sigma_1+\sin{\theta}\sigma_3,
\end{eqnarray}
with $\cos{\theta}=q/\sqrt{q^2+q_z^2}$ and $\mathbf{q}=(k,k_z)-\mathbf{L}$. Thus the Berry phase along this circle is $\nu\pi$, where $\nu=1$ is the winding number of $h(\mathbf{k})$ as $\theta$ varies from $0$ to $2\pi$.

\begin{figure}
\includegraphics[width=0.8\linewidth]{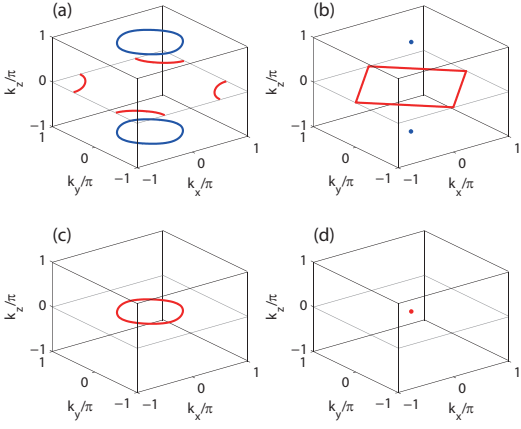}
\caption{Gap-closing regions with different parameters. (a) $m_0=0$, (b) $m_0=1$, (c) $m_0=2$ and (d) $m_0=3$.}\label{fig1}
\end{figure}

Generally speaking, the winding number $\nu$ can be any integer, and the corresponding Hamiltonian shall have the form
\begin{eqnarray}
h(\mathbf{k})=\cos{(\nu\theta)}\sigma_1+\sin{(\nu\theta)}\sigma_3.
\end{eqnarray}
Next we construct a model with $2\pi$-flux loops, the Hamiltonian of which reads:
\begin{eqnarray}
H_0&=&\sin{k_z}(\sum_{i=x,y,z}\cos{k_i}-m_0)\sigma_1\nonumber\\
&&+[\sin^2{k_z}-(\sum_{i=x,y,z}\cos{k_i}-m_0)^2]\sigma_3.\label{H}
\end{eqnarray}
Without loss of generality, hereafter we consider only the case with positive $m_0$.
The gap-closing condition is given by:
\begin{eqnarray}
&&M_1=\sin{k_z}=0,\nonumber\\
&&M_2=(\sum_{i=x,y,z}\cos{k_i}-m_0)=0.
\end{eqnarray}
Depending on the value of $m_0$, the system may hold different numbers of gap-closing loops, as illustrated in Fig.\ref{fig1}.
When $0<m_0<1$, the system holds two loops at $k_z=0$ and $k_z=\pi$, which center around $k_x=k_y=0$ and $k_x=k_y=\pi$ respectively, as shown by the blue and red lines in Fig.\ref{fig1}(a). The blue loop at $k_z=0$ will shrink into a point and disappear at $m_0=1$, while the red one will connect with itself at the edge of the Brillouin zone at $m_0=1$ [Fig.\ref{fig1}(b)], and transform into a loop centered around $k_x=k_y=0$ when $1<m_0<3$ [Fig.\ref{fig1}(c)]. This loop will also shrink into a point and disappear when $m_0=3$, as shown in Fig.\ref{fig1}(d).

\begin{figure}
\includegraphics[width=0.9\linewidth]{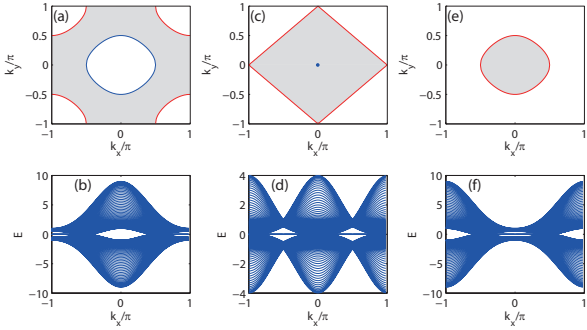}
\caption{Fourfold degenerate edge states under open boundary conditions along the $z$ direction. (a,b) $m_0=0$, (c,d) $m_0=1$ and (e,f) $m_0=2$. The edge states exist in the shadowed regions in (a), (c) and (e). In (b), (d) and (f) we demonstrate the energy spectrum with $k_y=0$. }\label{fig2}
\end{figure}

In Fig.\ref{fig2} we take open boundary condition along the $z$ direction, and demonstrate the edge states with different parameters. The edge states are fourfold degenerate in the shadowed regions, which correspond to a winding number of $2$.

\begin{figure}
\includegraphics[width=0.8\linewidth]{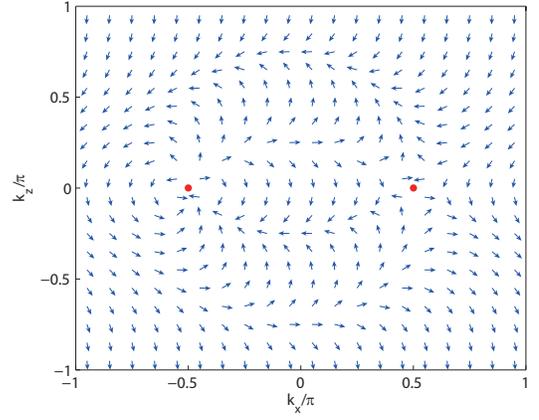}
\caption{The $2\pi$-flux loop and the pseudospin textures in the $k_x-k_z$ plane with $k_y=0$. The nodal loop is perpendicular to this plane, and is shown by the two red points.}\label{fig3}
\end{figure}

Next we calculate the winding number along a small trajectory enclosing the loop. Here we choose the single loop with $m_0=2$, while the winding number of the other loop at $m_0<1$ can also be calculated in the same way.
The Hamiltonian has quadratic dispersion near the loop as both $M_1$ and $M_2$ are linear in the small displacement $\mathbf{q}$.
Expanding the Hamiltonian in the $k_x-k_z$ plane as $k_x=\pi/2+q_x,~k_z=0+q_z,~k_y=0$,
we obtain
\begin{eqnarray}
h_1&=&\sin{k_z}\cos{k_x}\simeq -q_zq_x,\nonumber\\
h_3&=&[\sin^2{k_z}-\cos^2{k_x}]\simeq q_z^2-q_x^2.
\end{eqnarray}
By choosing
\begin{eqnarray}
q_x=A\cos{\theta},~q_z=A\sin{\theta},
\end{eqnarray}
the winding number for $\theta$ varying from $0$ to $2\pi$ is $\nu=2$, which corresponds to a $2\pi$ Berry flux of the loop. In Fig.\ref{fig3} we show the pseudospin textures in the $k_x-k_z$ plane with $k_y=0$. One can see that the pseudospin winds an angle of $2\pi$ along a trajectory enclosing each of the red points, which are the crossing points of the nodal loop and the plane.

\begin{figure*}
\includegraphics[width=0.8\linewidth]{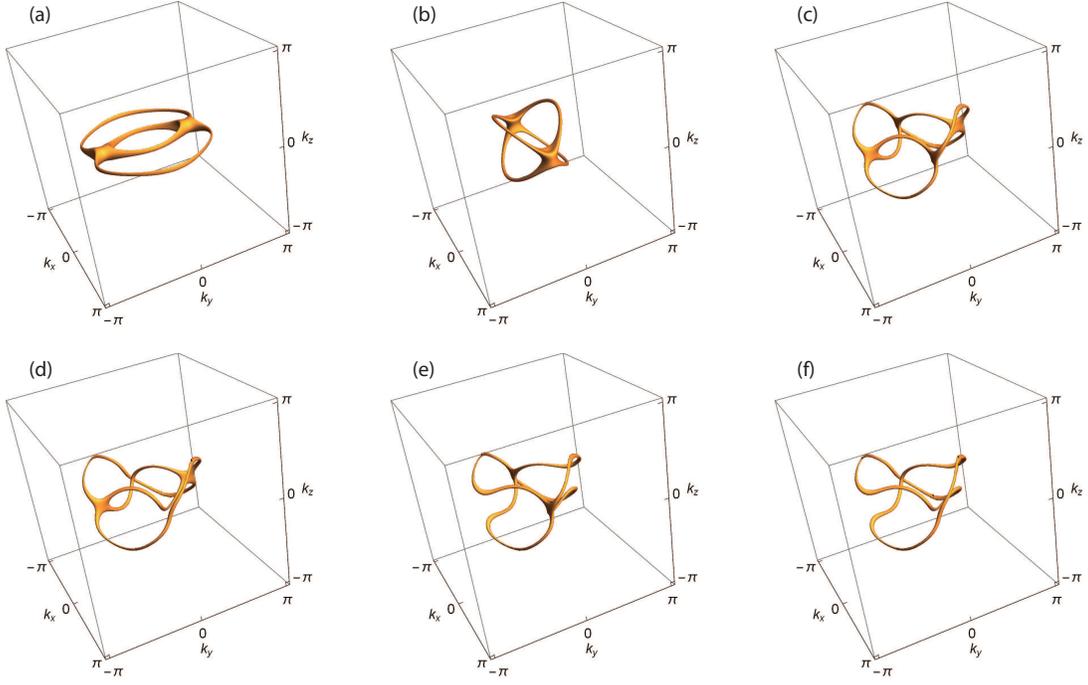}
\caption{Almost-zero energy surface of the Hamiltonian (\ref{total_H}) with $m_0=2$. (a) $A=C=0$, $B=1$; (b) $A=B=0$, $C=-1$; (c) $B=C=0$, $A=1$; (d) $A=1$, $B=0.2$, $C=0$; (e) $A=1$, $B=0$, $C=-0.2$ and (f) $A=1$, $B=0.2$, $C=-0.2$.}\label{fig4}
\end{figure*}

\emph{Hopf-link loops from a $2\pi$-flux loop.-}
A singularity with $\nu=2$ can also be seen as two degenerate singularities with $\nu=1$ each, and the degeneracy may be lifted by some perturbation. This fact holds also for a $2\pi$-flux loop, which may be decoupled into two loops by some extra terms. Specifically, if we choose the extra terms as
\begin{eqnarray}
H_1=A\sin{k_x}\sin{k_y}\sigma_1+(B\sin^2{k_x}+C\sin^2{k_y})\sigma_3,
\end{eqnarray}
the total Hamiltonian
\begin{eqnarray}
H=H_0+H_1\label{total_H}
\end{eqnarray}
gives the Hopf-link semimetal introduced in Ref.\cite{Yan_Hopf} when $A=B=-C=1$. By turning on each of the three terms, the $2\pi$-flux loop breaks into two loops with $\nu=1$ in different ways. In fig.\ref{fig4} we demonstrate the evolution of an almost zero-energy surface $E=\delta$, with $\delta\ll1$. A positive $B$ will shift the two loops to opposite directions along $k_x$ [Fig.\ref{fig4}(a)], a negative $C$ will rotate them around the $k_x$ axis [Fig.\ref{fig4}(b)], and a finite $A$ will break the degeneracy of the two loops in a more complicate manner [Fig.\ref{fig4}(c)]. The different contours of gapless regions suggest that the two types of loops ($2\pi$ flux and Hopf-link) may be related to certain symmetries, as discussed below.

\emph{Analysis of symmetries.-}
Generally speaking, gap-closing loops in a semimetal are protected by some symmetries of the Hamiltonian. In our model, the system preserves both a $P*T$ symmetry and a chiral symmetry $\Gamma=i\sigma_y$. Each of these two symmetries ensures the absence of a $\sigma_y$ term in the Hamiltonian\cite{Chen_Hopf}, and results in a 1D gap-closing region. However, neither of these symmetries distinguishes the difference between the $2\pi$-flux loop and the Hopf-link loops.

In order to find such symmetries, we begin with the single $2\pi$-flux loop in $k_z=0$ plane given by Hamiltonian (\ref{H}), which satisfies all the three reflection symmetries
\begin{eqnarray}
R_{\alpha}H_0(k_{\alpha})R_{\alpha}^{-1}=H_0(-k_{\alpha})
\end{eqnarray}
with $R_x=R_y=1$, $R_z=\sigma_z$. Here $R_x$ and $R_y$ reverse the loop and map it to itself, while $R_z$ maps the two regions divided by the $k_z=0$ plane, but keeps all the points of the loop unchanged. On the other hand, if a system holds two linked loops which do not touch each other, these loops must lay in different planes. In such a case, the system may still hold reflection symmetries with respect to the two planes which contain each of these two loops (i.e., when the two planes are perpendicular to each other), but the the third reflection symmetry must be broken.

Next, we consider Hamiltonian (\ref{total_H}) which describes a pair of Hopf-link loops. By introducing a finite $A$, the reflection symmetries $R_{\alpha}$ are broken in all three directions, and the system may support two linked nodal loops. However, the two loops are still degenerate at four points in the Brillouin zone, $k_z=k_i=0,~k_j=\pm\pi/2$ with $i,j=x,y$, as shown in Fig.\ref{fig4}(c).
Without changing the symmetries of the system, these crossing points can be removed with finite $B$ and $C$. By introducing a positive $B$, each point of $k_z=k_y=0,~k_x=\pm\pi/2$ will separate into two points along the $x$ axis in Fig.\ref{fig4}(d), while a negative $C$ separates each of $k_z=k_x=0,~k_y=\pm\pi/2$ along the $z$ axis in Fig.\ref{fig4}(e). Hence the coexistence of the B and C terms will lift the degeneracy on these four points, and results in two separated loops. It is also necessary that B and C have opposite signs, otherwise the two separated loops will not link to each other.

Finally, we note that although the reflection symmetries are broken, the system satisfies $C_2$ rotation symmetries around each of the $x$, $y$ and $z$ axis. The $C_2$ rotation symmetry around the $\alpha$ axis can be represented as
\begin{eqnarray}
C_{2,\alpha}H(k_{\beta},k_{\gamma})C_{2,\alpha}^{-1}=H(-k_{\beta},-k_{\gamma}),
\end{eqnarray}
with $\alpha\neq\beta\neq\gamma\neq\alpha$, and $C_{2,\alpha}=R_{\beta}R_{\gamma}$. Generally speaking, for a system with two equivalent loops (in the sense that they both carry a $\pi$ Berry phase), it is possible to find a transformation which maps the two loops into each other. In our model, this transformation is represented by either $C_{2,y}$ and $C_{2,z}$, while $C_{2,x}$ reverse each loop into itself.

\emph{Alternative realization of $2\pi$-flux loop semimetals.-}
The Hamiltonian Eq.~(\ref{H}) describes a $2\pi$-flux loop semimetal in a two-component lattice model, but in order to realize it, one needs to consider up to the fourth nearest-neighbour hopping terms, and the hopping parameters need to be fine tuned. Alternatively, it is possible to generate such a semimetal with only nearest-neighbour hopping terms, but with spin-orbit interaction. Consider a Hamiltonian describing a cubic lattice,
\begin{equation}\label{alternative_H}
H=\Delta\frac{1-\tau_z}{2}\frac{1-\delta s_z}{1-\delta^2}
+t\left(\sum_{i=x,y,z}\cos{k_i}-m_0 \right)\tau_x+\alpha\sin{k_z}\tau_x s_x.
\end{equation}
In this model $\tau_i$ and $s_i$ are the Pauli matrices acting on the two orbitals and spin-1/2 spaces, respectively, $t$ is the nearest neighbor hopping, and $\alpha$ is a spin-orbit coupling reminiscent of the Rashba spin-orbit interaction. The first term is a spin splitting for the upper orbital ($\tau_z=-1$) and its notation is chosen for later convenience, to simplify Eq.~(\ref{Heff}).

The Hamiltonian Eq.~(\ref{alternative_H}) describes a four-band system where the states with $\tau_z=1$ and $s_z=\pm1$ have zero energy when $\sum_{i=x,y,z}\cos{k_i}-m_0=\sin{k_z}=0$, i.e., on the same loops in the Brillouin zone of our original model. It is also clear that spin-conserving terms preserve the loop, thus we have omitted them. They would be present in general, but do not affect the presence and topological properties of the $2\pi$-flux loop. Numerical results also show that the Berry phase along a trajectory enclosing the loop is $\gamma=2\pi$.

These results, based on the full 4-band Hamiltonian, can be understood from a low-energy effective model valid when $\Delta(1\pm\delta)\gg t,\alpha$. We apply quasi-degenerate perturbation theory\cite{Winkler_book} taking the $\Delta$ term
as unperturbed Hamiltonian, and focus on the low-energy subspace with $\tau_z=1$. The perturbation is given by the off-diagonal $t$ and $\alpha$ terms.
By considering the perturbation up to the second-order and neglecting an overall constant, the final Hamiltonian is given by
\begin{eqnarray}\label{Heff}
H_{eff}&=&\frac{\delta\alpha^2}{\Delta}\left[\sin^2{k_z}-\left(\frac{t}{\alpha}\right)^2\left(\sum_{i=x,y,z}\cos{k_i}-m_0\right)^2\right]s_z\nonumber\\
&&-\frac{2t\alpha}{\Delta}\sin{k_z}\left(\sum_{i=x,y,z}\cos{k_i}-m_0\right)s_x,
\end{eqnarray}
which is equivalent to Eq.~(\ref{H}) except for the coefficients.

Finally, we note that $H_0$ and $H_1$ have similar structure and it is also possible to construct a lattice model of the Hopf-link semimetals in a similar way, by including an additional orbital.

\begin{figure}
\includegraphics[width=1\linewidth]{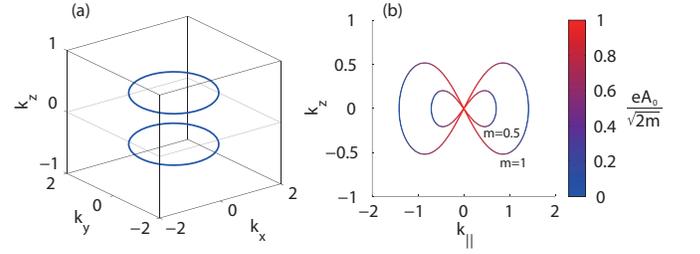}
\caption{The gap-closing region under a CPL along the $z$ direction. (a) the two loops induced by the CPL with $eA_0=0.5$ and $m=1$. (b) the motion of the loops with $eA_0$ varies from $0$ to $\sqrt{2m}$ with $m=0.5$ and $1$, $k_{\parallel}=\sqrt{k_x^2+k_y^2}$. When increasing $eA_0$, the loops will move in the $k$-space, and finally shrink into the origin when $eA_0=\sqrt{2m}$.
 }\label{fig5}
\end{figure}

\emph{Effect of periodic driving on $2\pi$-flux loop.-}
Next we consider a periodic drive generated by a circularly polarized light (CPL) propagating in the $z$ direction, which will divide the $2\pi$-flux loop into two separated loops with winding number $\nu=1$. The vector potential is given by $\mathbf{A}(t)=A_0(\cos{\omega t},\eta\sin{\omega t},0)$, where $\eta=1$ and $-1$ correspond to right-handed and left-handed CPL, respectively. The minimal coupling is given by $H(\mathbf{k})\rightarrow H(\mathbf{k}+e\mathbf{A}(t))$,  where we consider here the continuum limit of the lattice Hamiltonian (\ref{H}):
\begin{eqnarray}
H(\mathbf{k})=k_z(m-k^2/2)\sigma_1+[k_z^2-(m-k^2/2)^2]\sigma_3,\label{continuum}
\end{eqnarray}
with $m=3-m_0$.
The full Hamiltonian has a time-period $T=2\pi/\omega$, hence it can be expanded as $H(\mathbf{k},t)=\sum_n\mathcal{H}_n(\mathbf{k})e^{in\omega t}$, with
\begin{eqnarray}
\mathcal{H}_0(\mathbf{k})&=&k_z(\tilde{m}-k^2/2)\sigma_1\nonumber\\
&&+[k_z^2-(\tilde{m}-k^2/2)^2-e^2A_0^2(k_x^2+k_y^2)/2]\sigma_3,\nonumber\\
\mathcal{H}_{\pm1}(\mathbf{k})&=&-eA_0k_z(k_x\mp i\eta k_y)\sigma_1/2\nonumber\\
&&+eA_0(\tilde{m}-k^2/2)(k_x\mp i\eta k_y)\sigma_3,\nonumber\\
\mathcal{H}_{\pm2}(\mathbf{k})&=&-e^2A^2_0(k_x^2-k_y^2\mp2i\eta k_xk_y)\sigma_3/4,
\end{eqnarray}
where $\tilde{m}=m-e^2A_0^2/2$ and $\mathcal{H}_{n}(\mathbf{k})=0$ for $|n|>2$, despite the presence of $\sim k^4$ terms in Eq.(\ref{continuum}). In the limit where the driving
frequency $\omega$ is large comparing to other energy scales, the system can be described by an effective time-independent Hamiltonian\cite{Yan_2016,Yan_Hopf,Kitagawa_2011}:
\begin{eqnarray}
\mathcal{H}_{eff}(\mathbf{k})&=&\mathcal{H}_0+\sum_{n\geq1}\frac{[\mathcal{H}_{n},\mathcal{H}_{-n}]}{n\omega}+\mathcal{O}(\frac{1}{\omega})\nonumber\\
&=&k_z(\tilde{m}-k^2/2)\sigma_1\nonumber\\
&&+[k_z^2-(\tilde{m}-k^2/2)^2-e^2A_0^2(k_x^2+k_y^2)/2]\sigma_3.
\end{eqnarray}
The energy spectrum of $\mathcal{H}_{eff}(\mathbf{k})$ has two nodal loops given by the conditions
\begin{eqnarray}
\tilde{m}-k^2/2=0,~~~~~~
k_z^2-e^2A_0^2(k_x^2+k_y^2)/2=0.
\end{eqnarray}
These two loops are symmetric about $k_z=0$ (Fig.\ref{fig5}), and each of them has a winding number $\nu=1$. As we show in Fig.\ref{fig5}(b), $m$ does not affect qualitatively the evolution of the loops with drive strength, as long as $m$ is positive.

As a comparison, a lattice model describing conventional nodal loop semimetals is given by the Hamiltonian
\begin{eqnarray}
H=(\cos{k_x}+\cos{k_y}-m_0)\sigma_1+(m_1+\cos{k_z})\sigma_3,
\end{eqnarray}
which also holds a pair of nodal loops with $\pi$-flux each. However, these two loops have opposite winding number $\nu=\pm1$. By tuning the parameter $m_1$, these two loops will merge into each other and open a gap, instead of forming a $2\pi$-flux loop as discussed in this paper.
On the other hand, if the CPL propagates in an in-plane direction (e.g., along $x$ instead of $z$), it will deform the $2\pi$-flux loop into two pairs of Weyl points. These points will merge into each other and disappear with increasing $A_0$, which is similar to the case in Ref.\cite{Yan_2016}.

\emph{Conclusion.-}
We have introduced $2\pi$-flux loop semimetals which hold nodal loops carrying a Berry phase of $2\pi$. These loops can be described by a winding number $\nu=2$, and the system supports four-fold degenerate surface states under open boundary condition. The novel topological phase of Hopf-link semimetals can be generated from a $2\pi$-flux loop semimetal by breaking reflection symmetries while preserving $C_2$ rotation symmetries in three orthogonal directions. We then introduce a physical implementation of the $2\pi$-flux loops in a cubic lattice model with spin-orbit coupling. Finally, we study the Floquet properties of a $2\pi$-flux loop, and find that a circularly polarized light in the $z$ direction can drive the system into a nodal loop semimetal where the two $\pi$-flux loops have the same winding number $\nu=+1$.
On the other hand, light with circular polarization in the $x$ or $y$ direction will result in two pairs of Weyl points along $k_y=k_z=0$.

\emph{Acknowledgement.-}
Stefano Chesi acknowledges support from the National Key Research and Development
Program of China (Grant No. 2016YFA0301200) and NSFC (Grants No. 11574025 and No. U1530401).
Shu Chen is supported by the National Key Research and Development Program of China (2016YFA0300600), NSFC under Grants No. 11425419, No. 11374354 and No. 11174360, and the Strategic Priority Research Program (B) of the Chinese Academy of Sciences  (No. XDB07020000).

\bibliographystyle{apsrev4-1}
\bibliography{bibfile}

\begin{thebibliography}{29}%
\makeatletter
\providecommand \@ifxundefined [1]{%
 \@ifx{#1\undefined}
}%
\providecommand \@ifnum [1]{%
 \ifnum #1\expandafter \@firstoftwo
 \else \expandafter \@secondoftwo
 \fi
}%
\providecommand \@ifx [1]{%
 \ifx #1\expandafter \@firstoftwo
 \else \expandafter \@secondoftwo
 \fi
}%
\providecommand \natexlab [1]{#1}%
\providecommand \enquote  [1]{``#1''}%
\providecommand \bibnamefont  [1]{#1}%
\providecommand \bibfnamefont [1]{#1}%
\providecommand \citenamefont [1]{#1}%
\providecommand \href@noop [0]{\@secondoftwo}%
\providecommand \href [0]{\begingroup \@sanitize@url \@href}%
\providecommand \@href[1]{\@@startlink{#1}\@@href}%
\providecommand \@@href[1]{\endgroup#1\@@endlink}%
\providecommand \@sanitize@url [0]{\catcode `\\12\catcode `\$12\catcode
  `\&12\catcode `\#12\catcode `\^12\catcode `\_12\catcode `\%12\relax}%
\providecommand \@@startlink[1]{}%
\providecommand \@@endlink[0]{}%
\providecommand \url  [0]{\begingroup\@sanitize@url \@url }%
\providecommand \@url [1]{\endgroup\@href {#1}{\urlprefix }}%
\providecommand \urlprefix  [0]{URL }%
\providecommand \Eprint [0]{\href }%
\providecommand \doibase [0]{http://dx.doi.org/}%
\providecommand \selectlanguage [0]{\@gobble}%
\providecommand \bibinfo  [0]{\@secondoftwo}%
\providecommand \bibfield  [0]{\@secondoftwo}%
\providecommand \translation [1]{[#1]}%
\providecommand \BibitemOpen [0]{}%
\providecommand \bibitemStop [0]{}%
\providecommand \bibitemNoStop [0]{.\EOS\space}%
\providecommand \EOS [0]{\spacefactor3000\relax}%
\providecommand \BibitemShut  [1]{\csname bibitem#1\endcsname}%
\let\auto@bib@innerbib\@empty
\bibitem [{\citenamefont {Hasan}\ and\ \citenamefont
  {Kane}(2010)}]{Hasan_2010}%
  \BibitemOpen
  \bibfield  {author} {\bibinfo {author} {\bibfnamefont {M.~Z.}\ \bibnamefont
  {Hasan}}\ and\ \bibinfo {author} {\bibfnamefont {C.~L.}\ \bibnamefont
  {Kane}},\ }\href {\doibase 10.1103/revmodphys.82.3045} {\bibfield  {journal}
  {\bibinfo  {journal} {Rev. Mod. Phys.}\ }\textbf {\bibinfo {volume} {82}},\
  \bibinfo {pages} {3045} (\bibinfo {year} {2010})}\BibitemShut {NoStop}%
\bibitem [{\citenamefont {Qi}\ and\ \citenamefont {Zhang}(2011)}]{Qi_2011}%
  \BibitemOpen
  \bibfield  {author} {\bibinfo {author} {\bibfnamefont {X.-L.}\ \bibnamefont
  {Qi}}\ and\ \bibinfo {author} {\bibfnamefont {S.-C.}\ \bibnamefont {Zhang}},\
  }\href {\doibase 10.1103/revmodphys.83.1057} {\bibfield  {journal} {\bibinfo
  {journal} {Rev. Mod. Phys.}\ }\textbf {\bibinfo {volume} {83}},\ \bibinfo
  {pages} {1057} (\bibinfo {year} {2011})}\BibitemShut {NoStop}%
\bibitem [{\citenamefont {Shen}(2013)}]{Shen_book}%
  \BibitemOpen
  \bibfield  {author} {\bibinfo {author} {\bibfnamefont {S.-Q.}\ \bibnamefont
  {Shen}},\ }\href@noop {} {\emph {\bibinfo {title} {Topological Insulators}}}\
  (\bibinfo  {publisher} {Springer Berlin Heidelberg},\ \bibinfo {year}
  {2013})\BibitemShut {NoStop}%
\bibitem [{\citenamefont {Bernevig}\ and\ \citenamefont
  {Hughes}(2013)}]{Bernevig_book}%
  \BibitemOpen
  \bibfield  {author} {\bibinfo {author} {\bibfnamefont {B.~A.}\ \bibnamefont
  {Bernevig}}\ and\ \bibinfo {author} {\bibfnamefont {T.~L.}\ \bibnamefont
  {Hughes}},\ }\href@noop {} {\emph {\bibinfo {title} {Topological Insulators
  and Topological Superconductors}}}\ (\bibinfo  {publisher} {Princeton
  University Press},\ \bibinfo {year} {2013})\BibitemShut {NoStop}%
\bibitem [{\citenamefont {Wan}\ \emph {et~al.}(2011)\citenamefont {Wan},
  \citenamefont {Turner}, \citenamefont {Vishwanath},\ and\ \citenamefont
  {Savrasov}}]{Wan_2011}%
  \BibitemOpen
  \bibfield  {author} {\bibinfo {author} {\bibfnamefont {X.}~\bibnamefont
  {Wan}}, \bibinfo {author} {\bibfnamefont {A.~M.}\ \bibnamefont {Turner}},
  \bibinfo {author} {\bibfnamefont {A.}~\bibnamefont {Vishwanath}}, \ and\
  \bibinfo {author} {\bibfnamefont {S.~Y.}\ \bibnamefont {Savrasov}},\
  }\href@noop {} {\bibfield  {journal} {\bibinfo  {journal} {Phys. Rev. B}\
  }\textbf {\bibinfo {volume} {83}},\ \bibinfo {pages} {205101} (\bibinfo
  {year} {2011})}\BibitemShut {NoStop}%
\bibitem [{\citenamefont {Young}\ \emph {et~al.}(2012)\citenamefont {Young},
  \citenamefont {Zaheer}, \citenamefont {Teo}, \citenamefont {Kane},
  \citenamefont {Mele},\ and\ \citenamefont {Rappe}}]{Young_2012}%
  \BibitemOpen
  \bibfield  {author} {\bibinfo {author} {\bibfnamefont {S.~M.}\ \bibnamefont
  {Young}}, \bibinfo {author} {\bibfnamefont {S.}~\bibnamefont {Zaheer}},
  \bibinfo {author} {\bibfnamefont {J.~C.~Y.}\ \bibnamefont {Teo}}, \bibinfo
  {author} {\bibfnamefont {C.~L.}\ \bibnamefont {Kane}}, \bibinfo {author}
  {\bibfnamefont {E.~J.}\ \bibnamefont {Mele}}, \ and\ \bibinfo {author}
  {\bibfnamefont {A.~M.}\ \bibnamefont {Rappe}},\ }\href@noop {} {\bibfield
  {journal} {\bibinfo  {journal} {Phys. Rev. Lett.}\ }\textbf {\bibinfo
  {volume} {108}},\ \bibinfo {pages} {140405} (\bibinfo {year}
  {2012})}\BibitemShut {NoStop}%
\bibitem [{\citenamefont {Morimoto}\ and\ \citenamefont
  {Furusaki}(2014)}]{Morimoto_2014}%
  \BibitemOpen
  \bibfield  {author} {\bibinfo {author} {\bibfnamefont {T.}~\bibnamefont
  {Morimoto}}\ and\ \bibinfo {author} {\bibfnamefont {A.}~\bibnamefont
  {Furusaki}},\ }\href@noop {} {\bibfield  {journal} {\bibinfo  {journal}
  {Phys. Rev. B}\ }\textbf {\bibinfo {volume} {89}},\ \bibinfo {pages} {235127}
  (\bibinfo {year} {2014})}\BibitemShut {NoStop}%
\bibitem [{\citenamefont {Yang}\ and\ \citenamefont
  {Nagaosa}(2014)}]{Yang_2014}%
  \BibitemOpen
  \bibfield  {author} {\bibinfo {author} {\bibfnamefont {B.-J.}\ \bibnamefont
  {Yang}}\ and\ \bibinfo {author} {\bibfnamefont {N.}~\bibnamefont {Nagaosa}},\
  }\href {\doibase 10.1038/ncomms5898} {\bibfield  {journal} {\bibinfo
  {journal} {Nat. Commun.}\ }\textbf {\bibinfo {volume} {5}},\ \bibinfo {pages}
  {4898} (\bibinfo {year} {2014})}\BibitemShut {NoStop}%
\bibitem [{\citenamefont {Burkov}\ \emph {et~al.}(2011)\citenamefont {Burkov},
  \citenamefont {Hook},\ and\ \citenamefont {Balents}}]{Burkov_2011}%
  \BibitemOpen
  \bibfield  {author} {\bibinfo {author} {\bibfnamefont {A.~A.}\ \bibnamefont
  {Burkov}}, \bibinfo {author} {\bibfnamefont {M.~D.}\ \bibnamefont {Hook}}, \
  and\ \bibinfo {author} {\bibfnamefont {L.}~\bibnamefont {Balents}},\
  }\href@noop {} {\bibfield  {journal} {\bibinfo  {journal} {Phys. Rev. B}\
  }\textbf {\bibinfo {volume} {84}},\ \bibinfo {pages} {235126} (\bibinfo
  {year} {2011})}\BibitemShut {NoStop}%
\bibitem [{\citenamefont {Weng}\ \emph {et~al.}(2015)\citenamefont {Weng},
  \citenamefont {Liang}, \citenamefont {Xu}, \citenamefont {Yu}, \citenamefont
  {Fang}, \citenamefont {Dai},\ and\ \citenamefont {Kawazoe}}]{Weng_2015}%
  \BibitemOpen
  \bibfield  {author} {\bibinfo {author} {\bibfnamefont {H.}~\bibnamefont
  {Weng}}, \bibinfo {author} {\bibfnamefont {Y.}~\bibnamefont {Liang}},
  \bibinfo {author} {\bibfnamefont {Q.}~\bibnamefont {Xu}}, \bibinfo {author}
  {\bibfnamefont {R.}~\bibnamefont {Yu}}, \bibinfo {author} {\bibfnamefont
  {Z.}~\bibnamefont {Fang}}, \bibinfo {author} {\bibfnamefont {X.}~\bibnamefont
  {Dai}}, \ and\ \bibinfo {author} {\bibfnamefont {Y.}~\bibnamefont
  {Kawazoe}},\ }\href@noop {} {\bibfield  {journal} {\bibinfo  {journal} {Phys.
  Rev. B}\ }\textbf {\bibinfo {volume} {92}},\ \bibinfo {pages} {045108}
  (\bibinfo {year} {2015})}\BibitemShut {NoStop}%
\bibitem [{\citenamefont {Kim}\ \emph {et~al.}(2015)\citenamefont {Kim},
  \citenamefont {Wieder}, \citenamefont {Kane},\ and\ \citenamefont
  {Rappe}}]{Kim_2015}%
  \BibitemOpen
  \bibfield  {author} {\bibinfo {author} {\bibfnamefont {Y.}~\bibnamefont
  {Kim}}, \bibinfo {author} {\bibfnamefont {B.~J.}\ \bibnamefont {Wieder}},
  \bibinfo {author} {\bibfnamefont {C.}~\bibnamefont {Kane}}, \ and\ \bibinfo
  {author} {\bibfnamefont {A.~M.}\ \bibnamefont {Rappe}},\ }\href@noop {}
  {\bibfield  {journal} {\bibinfo  {journal} {Phys. Rev. Lett.}\ }\textbf
  {\bibinfo {volume} {115}},\ \bibinfo {pages} {036806} (\bibinfo {year}
  {2015})}\BibitemShut {NoStop}%
\bibitem [{\citenamefont {Yu}\ \emph {et~al.}(2015)\citenamefont {Yu},
  \citenamefont {Weng}, \citenamefont {Fang}, \citenamefont {Dai},\ and\
  \citenamefont {Hu}}]{Yu_2015}%
  \BibitemOpen
  \bibfield  {author} {\bibinfo {author} {\bibfnamefont {R.}~\bibnamefont
  {Yu}}, \bibinfo {author} {\bibfnamefont {H.}~\bibnamefont {Weng}}, \bibinfo
  {author} {\bibfnamefont {Z.}~\bibnamefont {Fang}}, \bibinfo {author}
  {\bibfnamefont {X.}~\bibnamefont {Dai}}, \ and\ \bibinfo {author}
  {\bibfnamefont {X.}~\bibnamefont {Hu}},\ }\href@noop {} {\bibfield  {journal}
  {\bibinfo  {journal} {Phys. Rev. Lett.}\ }\textbf {\bibinfo {volume} {115}},\
  \bibinfo {pages} {036807} (\bibinfo {year} {2015})}\BibitemShut {NoStop}%
\bibitem [{\citenamefont {Zhang}\ \emph {et~al.}(2016)\citenamefont {Zhang},
  \citenamefont {Zhao}, \citenamefont {Liu}, \citenamefont {Xue}, \citenamefont
  {Zhu},\ and\ \citenamefont {Wang}}]{Zhang_2016}%
  \BibitemOpen
  \bibfield  {author} {\bibinfo {author} {\bibfnamefont {D.-W.}\ \bibnamefont
  {Zhang}}, \bibinfo {author} {\bibfnamefont {Y.~X.}\ \bibnamefont {Zhao}},
  \bibinfo {author} {\bibfnamefont {R.-B.}\ \bibnamefont {Liu}}, \bibinfo
  {author} {\bibfnamefont {Z.-Y.}\ \bibnamefont {Xue}}, \bibinfo {author}
  {\bibfnamefont {S.-L.}\ \bibnamefont {Zhu}}, \ and\ \bibinfo {author}
  {\bibfnamefont {Z.~D.}\ \bibnamefont {Wang}},\ }\href@noop {} {\bibfield
  {journal} {\bibinfo  {journal} {Phys. Rev. A}\ }\textbf {\bibinfo {volume}
  {93}},\ \bibinfo {pages} {043617} (\bibinfo {year} {2016})}\BibitemShut
  {NoStop}%
\bibitem [{\citenamefont {Fang}\ \emph {et~al.}(2015)\citenamefont {Fang},
  \citenamefont {Chen}, \citenamefont {Kee},\ and\ \citenamefont
  {Fu}}]{Fang_2015}%
  \BibitemOpen
  \bibfield  {author} {\bibinfo {author} {\bibfnamefont {C.}~\bibnamefont
  {Fang}}, \bibinfo {author} {\bibfnamefont {Y.}~\bibnamefont {Chen}}, \bibinfo
  {author} {\bibfnamefont {H.-Y.}\ \bibnamefont {Kee}}, \ and\ \bibinfo
  {author} {\bibfnamefont {L.}~\bibnamefont {Fu}},\ }\href@noop {} {\bibfield
  {journal} {\bibinfo  {journal} {Phys. Rev. B}\ }\textbf {\bibinfo {volume}
  {92}},\ \bibinfo {pages} {081201} (\bibinfo {year} {2015})}\BibitemShut
  {NoStop}%
\bibitem [{\citenamefont {Fang}\ \emph {et~al.}(2016)\citenamefont {Fang},
  \citenamefont {Weng}, \citenamefont {Dai},\ and\ \citenamefont
  {Fang}}]{Fang_2016}%
  \BibitemOpen
  \bibfield  {author} {\bibinfo {author} {\bibfnamefont {C.}~\bibnamefont
  {Fang}}, \bibinfo {author} {\bibfnamefont {H.}~\bibnamefont {Weng}}, \bibinfo
  {author} {\bibfnamefont {X.}~\bibnamefont {Dai}}, \ and\ \bibinfo {author}
  {\bibfnamefont {Z.}~\bibnamefont {Fang}},\ }\href {\doibase
  10.1088/1674-1056/25/11/117106} {\bibfield  {journal} {\bibinfo  {journal}
  {Chin. Phys. B}\ }\textbf {\bibinfo {volume} {25}},\ \bibinfo {pages}
  {117106} (\bibinfo {year} {2016})}\BibitemShut {NoStop}%
\bibitem [{\citenamefont {Bzdu{\v{s}}ek}\ \emph {et~al.}(2016)\citenamefont
  {Bzdu{\v{s}}ek}, \citenamefont {Wu}, \citenamefont {Rüegg}, \citenamefont
  {Sigrist},\ and\ \citenamefont {Soluyanov}}]{Bzdusek_2016}%
  \BibitemOpen
  \bibfield  {author} {\bibinfo {author} {\bibfnamefont {T.}~\bibnamefont
  {Bzdu{\v{s}}ek}}, \bibinfo {author} {\bibfnamefont {Q.}~\bibnamefont {Wu}},
  \bibinfo {author} {\bibfnamefont {A.}~\bibnamefont {Rüegg}}, \bibinfo
  {author} {\bibfnamefont {M.}~\bibnamefont {Sigrist}}, \ and\ \bibinfo
  {author} {\bibfnamefont {A.~A.}\ \bibnamefont {Soluyanov}},\ }\href {\doibase
  10.1038/nature19099} {\bibfield  {journal} {\bibinfo  {journal} {Nature}\
  }\textbf {\bibinfo {volume} {538}},\ \bibinfo {pages} {75} (\bibinfo {year}
  {2016})}\BibitemShut {NoStop}%
\bibitem [{\citenamefont {Yu}\ \emph {et~al.}(2017)\citenamefont {Yu},
  \citenamefont {Wu}, \citenamefont {Fang},\ and\ \citenamefont
  {Weng}}]{Yu_2017}%
  \BibitemOpen
  \bibfield  {author} {\bibinfo {author} {\bibfnamefont {R.}~\bibnamefont
  {Yu}}, \bibinfo {author} {\bibfnamefont {Q.}~\bibnamefont {Wu}}, \bibinfo
  {author} {\bibfnamefont {Z.}~\bibnamefont {Fang}}, \ and\ \bibinfo {author}
  {\bibfnamefont {H.}~\bibnamefont {Weng}},\ }\href@noop {} {\bibfield
  {journal} {\bibinfo  {journal} {arXiv}\ } (\bibinfo {year} {2017})},\ \Eprint
  {http://arxiv.org/abs/1701.08502v1} {1701.08502v1} \BibitemShut {NoStop}%
\bibitem [{\citenamefont {Yan}\ and\ \citenamefont {Wang}(2016)}]{Yan_2016}%
  \BibitemOpen
  \bibfield  {author} {\bibinfo {author} {\bibfnamefont {Z.}~\bibnamefont
  {Yan}}\ and\ \bibinfo {author} {\bibfnamefont {Z.}~\bibnamefont {Wang}},\
  }\href@noop {} {\bibfield  {journal} {\bibinfo  {journal} {Phys. Rev. Lett.}\
  }\textbf {\bibinfo {volume} {117}},\ \bibinfo {pages} {087402} (\bibinfo
  {year} {2016})}\BibitemShut {NoStop}%
\bibitem [{\citenamefont {Lim}\ and\ \citenamefont
  {Moessner}(2017)}]{Lim_2017}%
  \BibitemOpen
  \bibfield  {author} {\bibinfo {author} {\bibfnamefont {L.-K.}\ \bibnamefont
  {Lim}}\ and\ \bibinfo {author} {\bibfnamefont {R.}~\bibnamefont {Moessner}},\
  }\href@noop {} {\bibfield  {journal} {\bibinfo  {journal} {Phys. Rev. Lett.}\
  }\textbf {\bibinfo {volume} {118}},\ \bibinfo {pages} {016401} (\bibinfo
  {year} {2017})}\BibitemShut {NoStop}%
\bibitem [{\citenamefont {Li}\ and\ \citenamefont
  {Ara{\'{u}}jo}(2016)}]{Li2016}%
  \BibitemOpen
  \bibfield  {author} {\bibinfo {author} {\bibfnamefont {L.}~\bibnamefont
  {Li}}\ and\ \bibinfo {author} {\bibfnamefont {M.~A.~N.}\ \bibnamefont
  {Ara{\'{u}}jo}},\ }\href {\doibase 10.1103/physrevb.94.165117} {\bibfield
  {journal} {\bibinfo  {journal} {Phys. Rev. B}\ }\textbf {\bibinfo {volume}
  {94}},\ \bibinfo {pages} {165117} (\bibinfo {year} {2016})}\BibitemShut
  {NoStop}%
\bibitem [{\citenamefont {Li}\ \emph {et~al.}(2017)\citenamefont {Li},
  \citenamefont {Yin}, \citenamefont {Chen},\ and\ \citenamefont
  {Ara{\'{u}}jo}}]{Li2017}%
  \BibitemOpen
  \bibfield  {author} {\bibinfo {author} {\bibfnamefont {L.}~\bibnamefont
  {Li}}, \bibinfo {author} {\bibfnamefont {C.}~\bibnamefont {Yin}}, \bibinfo
  {author} {\bibfnamefont {S.}~\bibnamefont {Chen}}, \ and\ \bibinfo {author}
  {\bibfnamefont {M.~A.~N.}\ \bibnamefont {Ara{\'{u}}jo}},\ }\href {\doibase
  10.1103/physrevb.95.121107} {\bibfield  {journal} {\bibinfo  {journal} {Phys.
  Rev. B}\ }\textbf {\bibinfo {volume} {95}},\ \bibinfo {pages} {121107}
  (\bibinfo {year} {2017})}\BibitemShut {NoStop}%
\bibitem [{\citenamefont {Chen}\ \emph {et~al.}(2017)\citenamefont {Chen},
  \citenamefont {Lu},\ and\ \citenamefont {Hou}}]{Chen_Hopf}%
  \BibitemOpen
  \bibfield  {author} {\bibinfo {author} {\bibfnamefont {W.}~\bibnamefont
  {Chen}}, \bibinfo {author} {\bibfnamefont {H.-Z.}\ \bibnamefont {Lu}}, \ and\
  \bibinfo {author} {\bibfnamefont {J.-M.}\ \bibnamefont {Hou}},\ }\href@noop
  {} {\bibfield  {journal} {\bibinfo  {journal} {arXiv}\ } (\bibinfo {year}
  {2017})},\ \Eprint {http://arxiv.org/abs/1703.10886v1} {1703.10886v1}
  \BibitemShut {NoStop}%
\bibitem [{\citenamefont {Yan}\ \emph {et~al.}(2017)\citenamefont {Yan},
  \citenamefont {Bi}, \citenamefont {Shen}, \citenamefont {Lu}, \citenamefont
  {Zhang},\ and\ \citenamefont {Wang}}]{Yan_Hopf}%
  \BibitemOpen
  \bibfield  {author} {\bibinfo {author} {\bibfnamefont {Z.}~\bibnamefont
  {Yan}}, \bibinfo {author} {\bibfnamefont {R.}~\bibnamefont {Bi}}, \bibinfo
  {author} {\bibfnamefont {H.}~\bibnamefont {Shen}}, \bibinfo {author}
  {\bibfnamefont {L.}~\bibnamefont {Lu}}, \bibinfo {author} {\bibfnamefont
  {S.-C.}\ \bibnamefont {Zhang}}, \ and\ \bibinfo {author} {\bibfnamefont
  {Z.}~\bibnamefont {Wang}},\ }\href@noop {} {\bibfield  {journal} {\bibinfo
  {journal} {arXiv}\ } (\bibinfo {year} {2017})},\ \Eprint
  {http://arxiv.org/abs/1704.00655v2} {1704.00655v2} \BibitemShut {NoStop}%
\bibitem [{\citenamefont {Chang}\ and\ \citenamefont {Yee}(2017)}]{Chang_Hopf}%
  \BibitemOpen
  \bibfield  {author} {\bibinfo {author} {\bibfnamefont {P.-Y.}\ \bibnamefont
  {Chang}}\ and\ \bibinfo {author} {\bibfnamefont {C.-H.}\ \bibnamefont
  {Yee}},\ }\href@noop {} {\bibfield  {journal} {\bibinfo  {journal} {arXiv}\ }
  (\bibinfo {year} {2017})},\ \Eprint {http://arxiv.org/abs/1704.01948v1}
  {1704.01948v1} \BibitemShut {NoStop}%
\bibitem [{\citenamefont {Ezawa}(2017)}]{Ezawa_Hopf}%
  \BibitemOpen
  \bibfield  {author} {\bibinfo {author} {\bibfnamefont {M.}~\bibnamefont
  {Ezawa}},\ }\href@noop {} {\bibfield  {journal} {\bibinfo  {journal} {arXiv}\
  } (\bibinfo {year} {2017})},\ \Eprint {http://arxiv.org/abs/1704.04941v1}
  {1704.04941v1} \BibitemShut {NoStop}%
\bibitem [{\citenamefont {Wilczek}\ and\ \citenamefont
  {Zee}(1983)}]{Wilczek_1983}%
  \BibitemOpen
  \bibfield  {author} {\bibinfo {author} {\bibfnamefont {F.}~\bibnamefont
  {Wilczek}}\ and\ \bibinfo {author} {\bibfnamefont {A.}~\bibnamefont {Zee}},\
  }\href {\doibase 10.1103/physrevlett.51.2250} {\bibfield  {journal} {\bibinfo
   {journal} {Phys. Rev. Lett.}\ }\textbf {\bibinfo {volume} {51}},\ \bibinfo
  {pages} {2250} (\bibinfo {year} {1983})}\BibitemShut {NoStop}%
\bibitem [{\citenamefont {Nakahara}(2003)}]{Nakahara_2003}%
  \BibitemOpen
  \bibfield  {author} {\bibinfo {author} {\bibfnamefont {M.}~\bibnamefont
  {Nakahara}},\ }\href
  {http://www.ebook.de/de/product/3646467/mikio_nakahara_geometry_topology_and_physics.html}
  {\emph {\bibinfo {title} {Geometry, Topology and Physics}}}\ (\bibinfo
  {publisher} {Taylor \& Francis Ltd},\ \bibinfo {year} {2003})\BibitemShut
  {NoStop}%
\bibitem [{\citenamefont {Winkler}(2003)}]{Winkler_book}%
  \BibitemOpen
  \bibfield  {author} {\bibinfo {author} {\bibfnamefont {R.}~\bibnamefont
  {Winkler}},\ }\href@noop {} {\emph {\bibinfo {title} {Spin-orbit Coupling
  Effects in Two-dimensional Electron and Hole Systems}}}\ (\bibinfo
  {publisher} {Springer-Verlag GmbH},\ \bibinfo {year} {2003})\BibitemShut
  {NoStop}%
\bibitem [{\citenamefont {Kitagawa}\ \emph {et~al.}(2011)\citenamefont
  {Kitagawa}, \citenamefont {Oka}, \citenamefont {Brataas}, \citenamefont
  {Fu},\ and\ \citenamefont {Demler}}]{Kitagawa_2011}%
  \BibitemOpen
  \bibfield  {author} {\bibinfo {author} {\bibfnamefont {T.}~\bibnamefont
  {Kitagawa}}, \bibinfo {author} {\bibfnamefont {T.}~\bibnamefont {Oka}},
  \bibinfo {author} {\bibfnamefont {A.}~\bibnamefont {Brataas}}, \bibinfo
  {author} {\bibfnamefont {L.}~\bibnamefont {Fu}}, \ and\ \bibinfo {author}
  {\bibfnamefont {E.}~\bibnamefont {Demler}},\ }\href {\doibase
  10.1103/physrevb.84.235108} {\bibfield  {journal} {\bibinfo  {journal} {Phys.
  Rev. B}\ }\textbf {\bibinfo {volume} {84}},\ \bibinfo {pages} {235108}
  (\bibinfo {year} {2011})}\BibitemShut {NoStop}%
\end{thebibliography}%

\end{document}